\documentclass[12pt,thmsa]{article}
\usepackage{amsfonts}

\usepackage{sw20aip}

\input{tcilatex}
\begin{document}

\title{Differential Geometry from Differential Equations}
\author{Simonetta Frittelli \\
%EndAName
Dept of Physics, Duquesne Univ., Pgh. PA15282 \\
Dept of Physics, Univ. of Pittsburgh, Pgh., PA15260 \and Carlos Kozameh \\
%EndAName
FaMAF, Universidad Nacional de Cordoba, 500 Cordoba, Argentina \and Ezra T.
Newman \\
%EndAName
Dept of Physics and Astronomy,\\
Univ. of Pittsburgh, Pittsburgh, PA 15260}
\date{April 27, 2001}
\maketitle

\begin{abstract}
{\small We first show how, from the general 3rd order ODE of the form }$%
z^{\prime \prime \prime }=F(z,z^{\prime },z^{\prime \prime },s)${\small ,
one can construct a natural Lorentzian conformal metric on the
four-dimensional space }$(z,z^{\prime },z^{\prime \prime },s).${\small \
When the function }$F(z,z^{\prime },z^{\prime \prime },s)${\small \
satisfies a special differential condition the conformal metric possesses a
conformal Killing field, }$\xi =\partial /\partial s,${\small \ which in
turn, allows the conformal metric to be mapped into a three dimensional
Lorentzian metric on the space }$(z,z^{\prime },z^{\prime \prime }${\small )
or equivalently, on the space of solutions of the original differential
equation. This construction is then generalized to the pair of differential
equations, }$z_{ss}=S(z,z_{s},z_{t},z_{st},s,t)${\small \ and }$%
z_{tt}=T(z,z_{s},z_{t},z_{st},s,t)${\small , with }$z_{s}${\small \ and }$%
z_{t}${\small \ the derivatives of }$z${\small \ with respect to }$s${\small %
\ and }$t${\small . In this case, from }$S${\small \ and }$T,${\small \ one
can again, in a natural manner, construct a Lorentzian conformal metric on
the six dimensional space }$(z,z_{s},z_{t},z_{st},s,t).${\small \ When the }$%
S${\small \ and }$T${\small \ satisfy differential conditions analogous to
those\ of the 3rd order ode,} {\small \ the 6-space then possesses a pair of
conformal Killing fields, }$\xi =\partial /\partial s${\small \ and }$\eta
=\partial /\partial t${\small \ which allows, via the mapping to the
four-space of (}$z,z_{s},z_{t},z_{st}${\small ) and a choice of conformal
factor, the construction of a four-dimensional Lorentzian metric. In fact
all four- dimensional Lorentzian metrics can be constructed in this manner.
This construction, with further conditions on }$S${\small \ and }$T,${\small %
\ thus includes all (local) solutions of the Einstein equations.}
\end{abstract}

\section{\protect\smallskip Introduction}

For the last 10 or so years we have been developing and studying a
reformulation of General Relativity where the primary objects of study are
families of 3-dimensional surfaces in a 4-space. They are used to define a
conformal Lorentzian metric via the requirement that the surfaces be null or
characteristic surfaces of that conformal metric; with, in addition, the
choice of a conformal factor to make the conformal metric into an Einstein
metric. From this point of view the metric tensor is a derived concept and
the Einstein equations appear as equations for the surfaces and the
conformal factor. In this study it had on occasion been useful to simplify
some of the equations by assuming we were studying conformal geometry in a
three-dimensional Lorentzian manifold. Much to our surprise, we recently
discovered that this 3-dimensional problem - with a totally different
motivation - had been studied in some classical papers by Cartan and Chern.
One major purpose of this work is to examine the relationship of our version
of 3-dimensional Lorentzian conformal geometry with that of Cartan and
Chern. Of perhaps greater importance is our generalization of these ideas to
the case of 4-dimensional conformal Lorentzian geometries. This, very much
more complicated, problem is discussed after an exposition of the
3-dimensional problem. Though one might consider these investigations to be
mainly in the realm of the study of certain classes of differential
equations, our main motivation has been towards the investigation of the
Einstein equations of general relativity. Already the theory of self- (or
anti-) dual vacuum Einstein metrics has arisen as a natural special case.
The emphasis now is on the full vacuum case.

In the late 1930's Cartan and Chern\cite{C1,C2,C3,Ch}, while studying the
invariance properties of differential equations, showed that there was a
natural geometric structure that can be associated with ordinary
differential equations [ODE's] of the form $z^{\prime \prime }=E(z,z^{\prime
},s)$ or $z^{\prime \prime \prime }=F(z,z^{\prime },z^{\prime \prime },s)$
where the prime denotes differentiation with respect to the independent
variable $s$. This geometric structure which is quite rich, involving a wide
variety of connections (projective, conformal and metric with and without
torsion), is given on the solution spaces of the equations.

More specifically, if the solutions are denoted by $z=z(x^{a},s)$ where $%
x^{a}$ are the arbitrary constants of integration, (two of them for the
first equation and three for the second) the geometric structures often (or
usually) live on the space of the constants of integration, but are
sometimes augmented with an extra dimension by adding the independent
variable $s$ as a fiber coordinate.

In Section II, we first review, from a new perspective, these results of
Cartan and Chern applied to the third order equation, and then, in Section
III, we generalize them to a new system of equations. For the new system we
will consider the dependent variable, $z$, to be a function of now \textit{%
two} independent variables, $s$ and $t,$ e.g., $z=z(s,t)$ that satisfies the
system of equations $z_{ss}=S(z,z_{s},z_{t},z_{st},s,t)$ and $%
z_{tt}=T(z,z_{s},z_{t},z_{st},s,t)$, with $z_{s}$ and $z_{t}$ the
derivatives of $z$ with respect to $s$ and $t$ . The integrability
conditions, $S_{tt}=T_{ss}$, are assumed to be satisfied. The solution space
to these equations, which is four-dimensional and is again augmented by the
fiber coordinates $s$ and $t,$ possesses a natural six-dimensional conformal
metric. By a judicious choice of the two functions, $S$ and $T$, (i.e. by
being solutions of a complicated differential equation) the six-dimensional
conformal metric possesses two conformal Killing fields and (via them and a
special choice of conformal factor) maps to a family of conformal Lorentzian
metrics on the four-dimensional solution space. All four-dimensional
Lorentzian metrics can be obtained in this manner. It follows that by a
further restriction in the choice of $S$ and $T$ and choice of conformal
factor, all Einstein spaces can be so obtained. In Section IV we discuss the
relationship of this work to the earlier work on the null surface
reformulation of GR.

For the sake of completeness, in Appendix A, we will outline the Cartan
geometry associated with the second order equation, $z^{\prime \prime
}=E(z,z^{\prime },s).$

\section{The Differential Geometry of $z^{\prime \prime \prime
}=F(z,z^{\prime },z^{\prime \prime },s)$}

We will study the geometry associated with the differential equation $
z^{\prime \prime \prime }=F(z,z^{\prime },z^{\prime \prime },s)$ assuming
that $F(z,z^{\prime },z^{\prime \prime },s)$ is a smooth function in all its
variables. We will only be interested in the local behavior of the
solutions. There will be several different (but related) spaces that will be
of interest to us. First of all, we mention the two-dimensional space of $%
(z,s);$ Cartan and Chern studied the problem of the equivalence classes of
differential equations under the diffeomorphisms in this two-space. This
problem, though of considerable interest, will not concern us. The next
space is the three-dimensional solution space of the differential equation.
The solutions of the third order ODE are given in terms of three constants
of integration, $x^{a},$ so that $z=z(x^{a},s)$ is the general solution and
the space of the $x^{a}$ is the solution space. For any fixed value of the
independent variable $s,$ the relations $\quad \quad $%
\begin{equation}
z=z(x^{a},s),\quad \quad z^{\prime }=z^{\prime }(x^{a},s),\quad \quad
z^{\prime \prime }=z^{\prime \prime }(x^{a},s)  \label{SOL}
\end{equation}
can be considered as a coordinate transformation (that depends on the
parameter $s),$ between the three $x^{a}$ and the ``coordinates'' $%
(z,z^{\prime },z^{\prime \prime }),$ i.e., it defines a one-parameter family
of coordinate transformations$.$ This leads naturally to the idea of a
four-dimensional space coordinatized either by ($x^{a},s)$ or by ($%
z,z^{\prime },z^{\prime \prime },s).$ The first choice suggests that this
four-space should be thought of as a base three-space augmented by the
one-dimensional fibers coordinatized by $s.$ Different constructions or
applications lead naturally to one or the other of the coordinatizations.

\begin{remark}
For most applications that are of interest to us, the independent variable $s
$ is taken to be the angle $\phi $ on the circle and the fiber would thus be
thought of as $S^{1}.$ This circle, in our applications to 3-dimensional
Lorentzian spaces, is simply the ``circle'' of null directions at each
space-time point. At this point in the exposition, this, however, is not
easily seen.\textbf{\ }Here, for the moment, we are only interested in the
local behavior.
\end{remark}

On the four dimensional space of ($z,z^{\prime },z^{\prime \prime },s),$ we
consider the four one-forms $\beta ^{a},$ 
\begin{eqnarray}
\beta ^{1} &=&dz-z^{\prime }ds,  \label{bforms} \\
\quad \beta ^{2} &=&dz^{\prime }-z^{\prime \prime }ds,  \nonumber \\
\quad \beta ^{3} &=&dz^{\prime \prime }-F(z,z^{\prime },z^{\prime \prime
},s)ds,  \nonumber \\
\quad \beta ^{4} &=&ds.  \nonumber
\end{eqnarray}
From (\ref{SOL}), we can write 
\begin{eqnarray*}
dz &=&\partial _{a}zdx^{a}+z^{\prime }ds, \\
dz^{\prime } &=&\partial _{a}z^{\prime }dx^{a}+z^{\prime \prime }ds, \\
dz^{\prime \prime } &=&\partial _{a}z^{\prime \prime }dx^{a}+F(z,z^{\prime
},z^{\prime \prime },s)ds,
\end{eqnarray*}
so that we have the alternative version of the forms

\begin{eqnarray}
\beta ^{1} &=&z_{a}dx^{a},  \label{bforms2} \\
\quad \beta ^{2} &=&z_{a}^{\prime }dx^{a},  \nonumber \\
\quad \beta ^{3} &=&z_{a}^{\prime \prime }dx^{a},  \nonumber \\
\quad \beta ^{4} &=&ds.  \nonumber
\end{eqnarray}

The following four linear combinations of the $\beta $'s will play a central
role, though for the moment only the first three will be used,

\begin{eqnarray}
\omega ^{1} &=&\beta ^{1},  \label{forms} \\
\omega ^{2} &=&\beta ^{2},  \nonumber \\
\omega ^{3} &=&\beta ^{3}+a\beta ^{1}+b\beta ^{2},  \nonumber \\
\omega ^{4} &=&C\beta ^{4}.  \nonumber
\end{eqnarray}
The ($a,b,C)$ are three functions of $(z,z^{\prime },z^{\prime \prime },s)$
that are to be determined.

From the $\omega ^{i}=$ ($\omega ^{1},\omega ^{2},\omega ^{3})$ we construct
the following one-parameter family of Lorentzian 3-metrics, parametrized by
the values of $s;$

\begin{equation}
g(z,z^{\prime },z^{\prime \prime },s)=\omega ^{1}\otimes \omega ^{3}+\omega
^{3}\otimes \omega ^{1}-\omega ^{2}\otimes \omega ^{2}.  \label{3metric}
\end{equation}
At this point we are simply defining a one-parameter family of metrics
constructed from the ($\omega ^{1},\omega ^{2},\omega ^{3}).$ Later we will
see that this definition is justified by the results.

Note: A more general version of (\ref{forms}) could have been used. The%
\textbf{\ }$\omega ^{2}$\textbf{\ }could have included another term so that
it had the form $\omega ^{2}=\beta ^{2}+A\beta ^{1}$ with the further
modification $\omega ^{3}=\beta ^{3}+(A+a)\beta ^{1}+(\frac{1}{2}%
A^{2}+b)\beta ^{2}.$ These exta terms however play no role; they form a
``null'' rotation, leaving the metric, (\ref{3metric}) invariant with
arbitrary $A.$)

\begin{remark}
In order to try to give some perspective and motivation we remark that from
another point of view, described in detail later, the metric (\ref{3metric})
arose from physical considerations where the function

$u=z(x^{a},s)=const.,$ represented a one-parameter family of null foliations
of a Lorentzian three-dimensional space-time. The three $\omega $'s are
chosen to form a null triad with $\omega ^{1}$ being the gradient of $%
z(x^{a},s)$ and with $\omega ^{3}$ as the second null covector of the triad.
The (so-far) arbitrary functions $a$ and $b$ will be \textit{uniquely}
determined by a requirement of ``minimal'' dependence of the metric, Eq.(\ref
{3metric}), on the parameter $s.$ The precise meaning of ` ``minimal''
dependence' will be given shortly.
\end{remark}

\begin{remark}
We emphasize that, though it appears that our choice of the form of the
metric Eq.(\ref{3metric}) is arbitrary, in fact, it appears to be the only
choice that allows the following construction. Later we will give, from the
other point of view, an alternate justification of its ``naturalness''.
\end{remark}

First we have:

\begin{definition}
For any function $H(z,z^{\prime },z^{\prime \prime },s),$ $\dot{H}$ is the
total $s-$derivative; 
\[
\frac{dH}{ds}\equiv \dot{H}\equiv H_{z}z^{\prime }+H_{z^{\prime }}z^{\prime
\prime }+H_{z^{\prime \prime }}F+H,_{s}.
\]
\end{definition}

Our plan is to first take the $s$-derivative of $g(z,z^{\prime },z^{\prime
\prime },s),$ i.e., 
\[
\dot{g}\equiv \frac{dg(z,z^{\prime },z^{\prime \prime },s)}{ds},\text{ } 
\]
and then by judicious choice of $a$ and $b,$ to make $\dot{g}$ \textit{as
close as possible} to being proportional to $g$ itself.

\begin{remark}
This derivative is actually the Lie derivative of the metric along the
vector field,\textbf{\ } 
\[
\frac{d}{ds}\equiv \frac{\partial }{\partial s}+z^{\prime }\frac{\partial }{%
\partial z}+z^{\prime \prime }\frac{\partial }{\partial z^{\prime }}+F\frac{%
\partial }{\partial z^{\prime \prime }},
\]

in the\textbf{\ }($z,z^{\prime },z^{\prime \prime },s)$ space$.$ It is,
perhaps, simpler to think of it as the total $s$-derivative of $g$\ in the%
\textbf{\ (}$z,z^{\prime },z^{\prime \prime },s)$\textbf{\ }coordinate
system.
\end{remark}

It turns out that to exactly make $\dot{g}=\lambda g$ requires a restriction
on the $F$ of the starting differential equation, $z^{\prime \prime \prime
}=F(z,z^{\prime },z^{\prime \prime },s).$ Nevertheless it is the ``\textit{%
as close as possible'' } condition that will constitute our `\thinspace
``minimal'' dependence' condition. Note that\textit{\ when }$\dot{g}=\lambda
g$ \textit{there is} a conformal structure naturally defined on the solution
space.

Explicitly the $s$-derivative of $g(z,z^{\prime },z^{\prime \prime },s),$ is

\begin{equation}
\dot{g}=\dot{\omega}^{1}\otimes \omega ^{3}+\omega ^{1}\otimes \dot{\omega}%
^{3}+\dot{\omega}^{3}\otimes \omega ^{1}+\omega ^{3}\otimes \dot{\omega}_{1}-%
\dot{\omega}^{2}\otimes \omega ^{2}-\omega ^{2}\otimes \dot{\omega}^{2}.
\label{gdot}
\end{equation}
From the definition of the forms we have for the first two $\omega $ that

\begin{eqnarray}
\dot{\omega}^{1} &=&\omega ^{2},  \label{omegadot} \\
\dot{\omega}^{2} &=&\omega ^{3}-a\omega ^{1}-b\omega ^{2},  \nonumber
\end{eqnarray}
Using

\begin{eqnarray*}
\dot{\omega}^{3} &=&\dot{\beta}^{3}+\dot{a}\omega ^{1}+a\dot{\omega}^{1}+%
\dot{b}\omega ^{2}+b\,\dot{\omega}^{2} \\
\dot{\beta}^{3} &=&z_{a}^{\prime \prime \prime
}dx^{a}=F,_{a}dx^{a}=F_{z}\omega ^{1}+F_{z^{\prime }}\omega
^{2}+F_{z^{\prime \prime }}(\omega ^{3}-a\omega ^{1}-b\omega ^{2})
\end{eqnarray*}
we have

\[
\dot{\omega}_{3}=(F_{z}-aF_{z^{\prime \prime }}+\dot{a}-ab)\omega
^{1}+(F_{z^{\prime }}-bF_{z^{\prime \prime }}+a+\dot{b}-b^{2})\omega
^{2}+(F_{z^{\prime \prime }}+b)\omega ^{3} 
\]
or

\begin{equation}
\dot{\omega}_{3}=U\omega ^{1}+V\omega ^{2}+W\omega ^{3},  \label{3dot}
\end{equation}
with

\begin{eqnarray}
U &=&F_{z}-aF_{z^{\prime \prime }}+\dot{a}-ab,  \label{UVW} \\
V &=&F_{z^{\prime }}-bF_{z^{\prime \prime }}+a+\dot{b}-b^{2},  \nonumber \\
W &=&F_{z^{\prime \prime }}+b.  \nonumber
\end{eqnarray}

Substituting Eqs.(\ref{omegadot}) and (\ref{3dot}) into Eq.(\ref{gdot}) we
obtain, after collecting terms,

\begin{equation}
\dot{g}=2U\omega ^{1}\otimes \omega ^{1}+2(V+a)\omega ^{(1}\otimes \omega
^{2)}+2W\omega ^{(1}\otimes \omega ^{3)}+2b\omega ^{2}\otimes \omega ^{2}.
\label{gdot2}
\end{equation}

We can now precisely state our condition of ``\textit{minimal} $s$ \textit{\
dependence}'' of the metric;

1. We require that, in Eq.(\ref{gdot2}), the coefficient of $\omega
^{(1}\otimes \omega ^{2)}$ vanishes, i.e., 
\begin{equation}
a=-V,  \label{c1}
\end{equation}

2. We require the two terms, $2W\omega ^{(1}\otimes \omega ^{3)}+2b\omega
^{2}\otimes \omega ^{2},$ combine so that they are proportional to the
metric, Eq.(\ref{3metric}), i.e., 
\begin{equation}
2b=-W.  \label{c2}
\end{equation}
This leads to the \textit{unique algebraic determination} of $a$ and $b$ in
terms of $F$ and its derivatives:

\begin{eqnarray}
b &=&-\frac{1}{3}F_{z^{\prime \prime }},  \label{ab} \\
2a &=&-F_{z^{\prime }}-\frac{2}{9}(F_{z^{\prime \prime }})^{2}+\frac{1}{3}%
\frac{d}{ds}(F_{z^{\prime \prime }}).  \nonumber
\end{eqnarray}
Using, from Eqs.(\ref{UVW}) and (\ref{ab}), 
\begin{equation}
U[F]\equiv F_{z}-a[F]F_{z^{\prime \prime }}+\dot{a}[F]-a[F]b[F],  \label{U}
\end{equation}
this leads to the final form of $\dot{g}:$%
\begin{equation}
\dot{g}[F]=2U[F]\omega ^{1}\otimes \omega ^{1}+\lambda g,  \label{gdot3}
\end{equation}
with 
\[
\lambda (x^{a},s)=\frac{2}{3}F_{z^{\prime \prime }}. 
\]
Our ``\textit{minimal} $s$ \textit{dependence}'' leads to a unique
determination of $a$ and $b$ and unique differential expressions for both $U$
and $\lambda $ in terms of $F.$

\begin{proposition}
Our one-parameter family of metrics are all conformally related if\textit{\ }%
$F$ \textit{is restricted by the condition }$U[F]=0,$ \textit{so that } 
\[
\dot{g}[F]=\frac{2}{3}F_{z^{\prime \prime }}g.
\]
In this case there exists a conformal factor, $\Omega ,$ with $\dot{\Omega }=%
\frac{1}{3}F_{z^{\prime \prime }}\Omega ,$ so that, for all values of $s,$%
\textbf{\ }the metric\textbf{\ }$\widehat{g}=$ $\Omega ^{-2}g,$ satisfies $%
\frac{d}{ds}\widehat{g}=0$.
\end{proposition}

In the general case, $U[F]\neq 0,$ we can extend the metric $g$ to a four
dimensional metric by 
\begin{equation}
g^{(4)}=g-\omega ^{4}\otimes \omega ^{4}=g-C^{2}ds\otimes ds  \label{g4}
\end{equation}
so that 
\begin{eqnarray}
\dot{g}^{(4)} &=&\dot{g}-2C\dot{C}ds\otimes ds,  \label{g4dot} \\
&=&2U\omega ^{1}\otimes \omega ^{1}+\frac{2}{3}F_{z^{\prime \prime }}g-2C%
\dot{C}ds\otimes ds,  \nonumber
\end{eqnarray}
If the unknown $C$ is chosen such that $\dot{C}=\frac{1}{3}F_{z^{\prime
\prime }}C$ then 
\begin{equation}
\dot{g}^{(4)}=2U\omega ^{1}\otimes \omega ^{1}+\frac{2}{3}F_{z^{\prime
\prime }}g^{(4)}.  \nonumber
\end{equation}

\begin{proposition}
For the special case of $U[F]=0,$ we see that $\xi =d/ds$ is a conformal
Killing field of the four-space so that each of the ``three-slices'', $%
s=constant,$ yield three-metrics that are conformally related. If $g^{(4)}$
is conformally rescaled by $\widehat{g}^{(4)}=$ $\Omega ^{-2}g^{(4)}$, with $%
\dot{\Omega }=\frac{1}{3}F_{z^{\prime \prime }}\Omega ,$ the conformal
Killing vector field becomes a Killing field and the three-slices are all
isometric.
\end{proposition}

An alternate point of view\cite{C1} towards the geometry of $z^{\prime
\prime \prime }=F(z,z^{\prime },z^{\prime \prime },s)$ is via the first of
Cartan's structure equations, for the three one-forms 
\[
\omega ^{i}=(\omega ^{1},\omega ^{2},\omega ^{3}) 
\]
we have$;$

%\begin{mathletters}
\begin{equation}
d\omega ^{i}=\omega _{\,j}^{i}\wedge \omega ^{j}+T^{i}.  \label{Structure}
\end{equation}
The indices are raised and lowered with the Lorentzian metric, (from Eq.(\ref
{3metric})), $\eta _{ij,}$ with $\eta _{13}$ = -$\eta _{22},$ all other
independent components vanishing. The basis one-forms are taken as $dx^{a}$
and $ds;$ so that though the $\omega ^{i}$ contain only $dx^{a},$ the other
forms, i.e., $d\omega ^{i}$, $\omega _{j}^{i}$ and $T^{i}$ will, in general,
contain $ds.$ The connection one-forms, $\omega _{\text{\thinspace
\thinspace }j}^{i},$ do not form a metric connection but rather a conformal
connection. Written as $\omega _{ij}=\eta _{ik}\omega _{\text{\thinspace
\thinspace }j}^{k}$ they are given by

%\end{mathletters}
\begin{eqnarray}
\omega _{ij} &=&w_{ij}+\widehat{\omega }\eta _{ij},  \label{connection} \\
w_{ij} &=&-w_{ji}
\end{eqnarray}
i.e., are taken as a metric connection plus a trace-term, ($\widehat{\omega }%
=\frac{1}{3}\omega _{i}^{i}).$

\begin{remark}
Note that the use of the trace term in the connection, $\omega _{ij},$ is a
variant of a (Weyl) connection via the equation\textbf{\ }$\nabla
_{c}g_{ab}(x^{a})=2\widehat{\omega }_{c}g_{ab}.$ It is not exactly the same
as a Weyl connection, but is a variant of it, because here we have the extra
degree of freedom, namely the variable $s$.
\end{remark}

Writing out Eq.(\ref{Structure}) we have

\begin{eqnarray}
d\omega ^{1} &=&(w_{[31]}+\widehat{\omega })\wedge \omega
^{1}+w_{[32]}\wedge \omega ^{2}+T^{1},  \label{structure2} \\
d\omega ^{2} &=&-w_{[21]}\wedge \omega ^{1}+\widehat{\omega }\wedge \omega
^{2}-w_{[23]}\wedge \omega ^{3}+T^{2},  \nonumber \\
d\omega ^{3} &=&w_{[12]}\wedge \omega ^{2}+(w_{[13]}+\widehat{\omega }
)\wedge \omega ^{3}+T^{3}  \nonumber
\end{eqnarray}
which, to determine the structure and torsion forms, can be compared with
the direct calculation of $d\omega ^{i},$ namely

\begin{eqnarray}
d\omega ^{1} &=&ds\wedge \omega ^{2},  \label{strut} \\
d\omega ^{2} &=&ds\wedge (\omega ^{3}-a\omega ^{1}-b\omega ^{2}),  \nonumber
\\
d\omega ^{3} &=&ds\wedge (\omega ^{1}U[F]-a\omega ^{1}-2b\omega ^{2}) 
\nonumber \\
&&+(a_{z^{\prime }}-ba_{z^{\prime \prime }}-b_{z}+ab_{z^{\prime \prime
}})\omega ^{2}\wedge \omega ^{1}+a_{z^{\prime \prime }}\omega ^{3}\wedge
\omega ^{1}+b_{z^{\prime \prime }}\omega ^{3}\wedge \omega ^{2}.  \nonumber
\end{eqnarray}
When the comparison is made we see that there are far more variables than
equations and thus there are ambiguities in the algebraic solution for $%
w_{ij},$ $\widehat{\omega }$ and $T^{i}.$ If however we require that the
skew-part of the connection, when pulled back to the constant $s$ surfaces,
be precisely the \textit{metric connection }of\textit{\ }Eq.\textit{(\ref
{3metric}), }then we have a unique solution for the connection and torsion:

\begin{eqnarray}
w_{[32]} &=&ds-\frac{1}{2}b_{z^{\prime \prime }}\omega ^{1},
\label{solution} \\
w_{[31]} &=&-a_{z^{\prime \prime }}\omega ^{1}-\frac{1}{2}b_{z^{\prime
\prime }}\omega ^{2}+bds,  \nonumber \\
w_{[12]} &=&-ads+(a_{z^{\prime }}-ba_{z^{\prime \prime
}}-b_{z}+ab_{z^{\prime \prime }})\omega ^{1}-\frac{1}{2}b_{z^{\prime \prime
}}\omega ^{3},  \nonumber \\
\widehat{\omega } &=&-bds,  \nonumber \\
T^{1} &=&0,\text{ }\qquad \text{ }T^{2}=0,\qquad T^{3}=U[F]ds\wedge \omega
^{1}.  \nonumber
\end{eqnarray}
Once again we see the geometric role of $U[F];$ when it vanishes the
``conformal'' connection has zero torsion.

We thus have seen that the general third order differential equation

\[
z^{\prime \prime \prime }=F(z,z^{\prime },z^{\prime \prime },s) 
\]
induces a variety of geometric structures; a ``conformal'' connection on the
solutions space, $x^{a},$ a four-dimensional Lorentzian metric on the space (%
$z,z^{\prime },z^{\prime \prime },s)$ so that when the space is foliated by
the constant $s$ three-surfaces they possess a one parameter family of
three-metrics, all closely related, satisfying $\dot{g}=2U[F]\omega
^{1}\otimes \omega ^{1}+\frac{2}{3}F_{z^{\prime \prime }}g.$ When the
special condition $U[F]=0$ is satisfied all the three-metrics are
conformally equivalent. Cartan studied the connection associated with the
full conformal equivalence class. We, instead, worked out the metric
connections, Eq.(\ref{solution}), associated with the one-parameter family
of metrics, Eq.(\ref{3metric}).

\begin{remark}
The study\cite{C1,C2,C3,Ch,W} of this third order ODE had its origin in the
classical question of the equivalence of ODE's under transformations in the
plane; $(z,s)\Leftrightarrow (z^{*},s^{*}).$ Cartan studied the equivalence
classes (with their invariants) of 3rd order ODEs under point transformations%
\textbf{, }$z^{*}=\frak{Z}(z,s),$\textbf{\ }$s^{*}=\frak{S}(z,s),$\textbf{\ }%
while Chern studied the same problems but under a larger group of
transformation, the group of contact transformations. The functional $U[F],$
often referred as the Wunschmann Invariant, is a relative invariant under
contact transformations of the 3rd order ODEs. We return briefly to this
issue in the Discussion section.
\end{remark}

To conclude this section we will summarize\cite{CK,T}, what appears to be a
completely different problem that in fact turns out to be virtually
identical, or more correctly, turns out to be the inverse\textbf{\ }to the
problem just addressed, namely the geometry of 3rd order ODEs. Roughly
speaking, we begin with 3-dimensional conformal Lorentzian metric and find a
complete integral of the associated Eikonal equation; i.e., we find a one
parameter,\textbf{\ }$\mathbf{`s}$', family of characteristic surfaces of
sufficient generality. By taking three derivatives with respect to the
parameter, the three space-time coordinates can be eliminated from the
eikonal, resulting in a 3rd order ODE with $\mathbf{`s}$'\ being the
independent variable. Automatically the Wunschmann Invariant vanishes.
Actually, (along with the problem of the four-dimensional solution space of
the next section) this inverse point of view was how we first addressed the
issues of this work.

More precisely, we begin with a three manifold, $\frak{M,}$ locally
coordinatized by, $x^{a},$ and require that there be a Lorentzian
(conformal) metric determined in the following manner: there is to exist a
one-parameter family of foliations of $\frak{M,}$ of sufficient generality,
(referred to as a complete integral) such that every member of the foliation
is to be a null-surface of the unknown metric. If the level surfaces of the
one-parameter family of foliations (parametrized by $s$) is given by

\[
u=z(x^{a},s), 
\]
then the condition that they be null, for all values of $s$, with respect to
the unknown metric, $g^{ab}(x^{a}),$ is

\begin{equation}
g^{ab}\partial _{a}z\partial _{b}z\equiv g^{ab}z_{a}z_{b}=0.  \label{null}
\end{equation}

By now taking a series (four) of $s$ derivatives of Eq.(\ref{null}), we have

\begin{eqnarray}
g^{ab}z_{a}^{\prime }z_{b} &=&0,  \label{components} \\
g^{ab}z_{a}^{\prime \prime }z_{b}+g^{ab}z_{a}^{\prime }z_{b}^{\prime } &=&0,
\label{components2} \\
g^{ab}z_{a}^{\prime \prime \prime }z_{b}+3g^{ab}z_{a}^{\prime \prime
}z_{b}^{\prime } &=&0,  \label{components3} \\
g^{ab}z_{a}^{\prime \prime \prime \prime }z_{b}+4g^{ab}z_{a}^{\prime \prime
\prime }z_{b}^{\prime }+3g^{ab}z_{a}^{\prime \prime }z_{b}^{\prime \prime }
&=&0.  \label{components4}
\end{eqnarray}

Then by considering the set

\begin{equation}
u=z(x^{a},s),\quad w=z^{\prime }(x^{a},s),\quad R=z^{\prime \prime
}(x^{a},s),\quad F=z^{\prime \prime \prime }(x^{a},s)  \label{inverting}
\end{equation}
the three $x^{a}$ can be eliminated from the last expression via the first
three expressions, yielding

\begin{equation}
z^{\prime \prime \prime }=F(z,z^{\prime },z^{\prime \prime },s).  \label{'''}
\end{equation}
From the form of the metric, (\ref{3metric}), we have that $\omega
^{1}=z_{a}dx^{a}$ \textbf{\ }is a null covector. This observation thus
establishes the connection with the first approach.

[Note that by ``a one parameter family of sufficient generality'', we mean
that the three one-forms\textbf{\ (}$z_{a}dx^{a},z_{a}^{\prime }dx^{a},$ $%
z_{a}^{\prime \prime }dx^{a})$ are linearly independent for all\textbf{\ }$%
\mathbf{s}$\textbf{. }This implies that one can invert (\ref{inverting}),
i.e., obtain\textbf{\ }$x^{a}=X^{a}(z,z^{\prime },z^{\prime \prime },s).$]

We see that, via Eq.(\ref{'''}), $z_{a}^{\prime \prime \prime }$ and $%
z_{a}^{\prime \prime \prime \prime }$ can be expressed in terms of the
gradient basis ($z_{a},z_{a}^{\prime },z_{a}^{\prime \prime }).$ The five
expressions Eqs.(\ref{null}) and (\ref{components}) yield the five
independent components of a conformal metric that depends on $s.$ This
conformal metric $($though described in the gradient basis) is identical to
the conformal metric of Eq.(\ref{3metric}) which is described in a null
basis.

The fifth derivative of Eq.(\ref{null}), namely

\[
g^{ab}z_{a}^{(5)}z_{b}+5g^{ab}z_{a}^{(4)}z_{b}^{\prime
}+10g^{ab}z_{a}^{\prime \prime \prime }z_{b}^{\prime \prime }=0, 
\]
when expressed in terms of $F$ and its derivatives, is identical to Eq.(\ref
{U}), i.e., $U[F]=0.$ This completes the display of the equivalence of the
two approaches. It also gives the justification for the apparently arbitrary
choices of the one-forms (\ref{forms}) and metric (\ref{3metric}). Linear
combinations of the three gradient one-forms,\textbf{\ (}$%
z_{a}dx^{a},z_{a}^{\prime }dx^{a},$ $z_{a}^{\prime \prime }dx^{a}),$\textbf{
\ }form the null triad \textbf{(}$\omega ^{1},\omega ^{2},\omega ^{3})$%
\textbf{. }

In addition to the condition $U[F]=0,$ Tod\cite{Tod}, following Cartan\cite
{C2}, in a continuing study of the Eq.(\ref{'''}) imposes further
restrictions on $F\ $so that the resulting metrics contain all
three-dimensional Einstein -Weyl spaces.

\section{Pairs of Partial Differential Equations}

The discussion of the previous section is really a variant of the work of
Cartan and Chern with our point of view. In this section we will discuss a
new situation. We want to find differential equations whose solution space
is four-dimensional and in addition possess a Lorentzian structure. This
four-dimensional solution space is to be the four-dimensional manifold $%
\frak{M}$ of physical space-time. Our goal, eventually, is to impose the
Einstein vacuum equations on this space. This issue however will not be
addressed here\textbf{.} After the consideration of the equation $z^{\prime
\prime \prime }=F(z,z^{\prime },z^{\prime \prime },s)$ one might have
thought that the generalization\textbf{\ }from three to four dimensions
should be to an equation of the form, $z^{\prime \prime \prime \prime
}=G(z,z^{\prime },z^{\prime \prime },z^{\prime \prime \prime },s)$ whose
solution space \textit{is four dimensional. }This case was studied by Bryant%
\cite{B} who found a further rich variety of geometric structures, e.g., a
quartic metric, $g_{abcd}$ but \textbf{it} does not include a
four-dimensional Lorentzian structure.

We have taken a different direction for the creation of a four-dimensional
solution space; we consider and study the geometry of the pair of equations

\begin{equation}
Z_{ss}=P(Z,Z_{s},Z_{t},Z_{st},s,t),\;Z_{tt}=Q(Z,Z_{s},Z_{t},Z_{st},s,t),
\label{pairs}
\end{equation}
where $P$ and $Q$ satisfy the integrability conditions for all $Z$,

\begin{equation}
D_{t}^{2}P=D_{s}^{2}Q  \label{Int}
\end{equation}
and the weak inequality, needed for the four-dimensionality of the solution
space,

\begin{equation}
1>(\frac{\partial P}{\partial Z_{ts}})(\frac{\partial Q}{\partial Z_{ts}}).
\label{Ineq}
\end{equation}

We have used the notation for the total derivatives $D_{t}$ or $D_{s}$ to
mean, respectively, the $t$ and $s$ derivatives acting on all the variables
but holding, respectively, the $s$ or the $t$ constant. For example, if $%
H=H(Z,Z_{s},Z_{t},Z_{st},s,t)$ then

\begin{equation}
D_{t}H\equiv \frac{\partial H}{\partial Z}Z_{t}+\frac{\partial H}{\partial %
Z_{s}}Z_{ts}+\frac{\partial H}{\partial Z_{t}}Z_{tt}+\frac{\partial H}{%
\partial Z_{ts}}Z_{tts}+\frac{\partial H}{\partial t}.  \label{totalD}
\end{equation}
$D_{t}$ and $D_{s}$ should also be thought of as vector fields on the
six-space, $(Z,Z_{s},Z_{t},Z_{st},s,t).$

For notational reasons and for comparison with earlier work but without
changing anything essential, we will consider $P$ and $Q$ to be complex
conjugates of each other, $s$ and $t$ to also be complex conjugates of each
other and will adopt the notation that $P=S\ $and $Q=S^{*}$ and $t=s^{*}$
with $D_{s}\equiv D$ and $D_{t}\equiv D^{*}.$ (For example, $s$ and $s^{*}$
can be considered as the complex stereographic coordinates on $S^{2}.)$

The solution space of Eqs.(\ref{pairs}) is four-dimensional\cite{N}, the
space of constants of integration, $(x^{a});$ solutions can be written as

\[
Z=Z(x^{a},s,s^{*}). 
\]
We will be interested in several different spaces; the four dimensional
space of the $(x^{a});$ the space of 
\begin{equation}
(Z,Z_{s},Z_{s^{*}},Z_{ss^{*}})\equiv (Z,DZ,D^{*}Z,DD^{*}Z)\equiv
(Z,W,W^{*},R),  \label{define}
\end{equation}
(defining the $Z,W,W^{*},R)$ and the six-dimensional space of ($
Z,W,W^{*},R,s,s^{*}).$

Our starting equations are then rewritten

\begin{eqnarray}
D^{2}Z &=&S(Z,DZ,D^{*}Z,DD^{*}Z,s,s^{*}),  \label{pair2} \\
D^{*2}Z &=&S^{*}(Z,DZ,D^{*}Z,DD^{*}Z,s,s^{*}).  \nonumber
\end{eqnarray}

We identify the spaces 
\begin{equation}
(x^{a})\Leftrightarrow (Z,W,W^{*},R)  \label{transf}
\end{equation}
for any fixed values of ($s,s^{*}),$ treating the relationship, Eq.(\ref
{transf}), as a coordinate transformation between the two sets, that is
parametrized by ($s,s^{*}).$ The six-space can then be coordinatized either
by ($x^{a},s,s^{*})$ or by ($Z,W,W^{*},R,s,s^{*})$. It is useful to think of
the larger space as being a two-dimensional bundle over the four-space, $%
x^{a}.$ In our applications it is taken to be the sphere-bundle, physically,
the bundle of null directions at each space-time point. This point of view
will not be emphasized here.

We begin with the six gradient one-forms 
\[
\theta ^{i}=(\theta ^{0},\theta ^{+},\theta ^{-},\theta ^{1})\equiv \partial
_{a}(Z,W,W^{*},R)dx^{a} 
\]

\begin{eqnarray}
\theta ^{0} &\equiv &dZ-Wds-W^{*}ds^{*}=Z_{a}dx^{a},  \label{betas} \\
\quad \theta ^{+} &\equiv &dW-D^{2}Zds-DD^{*}Zds^{*}=W_{a}dx^{a},  \nonumber
\\
\quad \theta ^{-} &\equiv &dW^{*}-Rds-D^{*2}Zds^{*}=W_{a}^{*}dx^{a}, 
\nonumber \\
\quad \theta ^{1} &\equiv &dR-D^{*}D^{2}Zds-DD^{*2}ds^{*}=R_{a}dx^{a}, 
\nonumber
\end{eqnarray}

\begin{eqnarray}
\theta &\equiv &ds, \\
\theta ^{*} &\equiv &ds^{*},  \nonumber
\end{eqnarray}
and form the combinations $\omega ^{i}=(\omega ^{0},\omega ^{+},\omega
^{-},\omega ^{1})$

\begin{eqnarray}
\omega ^{0} &=&\theta ^{0},  \label{one-forms} \\
\omega ^{+} &=&\alpha (\theta ^{+}+b\theta ^{-}),  \nonumber \\
\omega ^{-} &=&\alpha (\theta ^{-}+b^{*}\theta ^{+}),  \nonumber \\
\omega ^{1} &=&(\theta ^{1}+a\theta ^{+}+a^{*}\theta ^{-}+c\theta ^{0}), 
\nonumber
\end{eqnarray}
and

\begin{eqnarray}
\omega &=&C\theta , \\
\omega ^{*} &=&C^{*}\theta ^{*}  \nonumber
\end{eqnarray}
where the ($\alpha ,a,b,c)$ and $C$ are to be determined.

\subsection{Four-Dimensional Lorentzian Metrics}

From the four $\omega ^{i},$ we form the 2-parameter, ($s,s^{*})$ family, of
Lorentzian four-metrics by

\begin{eqnarray}
g(x^{a},s,s^{*}) &=&\omega ^{0}\otimes \omega ^{1}+\omega ^{1}\otimes \omega
^{0}-\omega ^{+}\otimes \omega ^{-}-\omega ^{-}\otimes \omega ^{+},
\label{4metric} \\
&=&\eta _{ij}\omega ^{i}\otimes \omega ^{j}  \nonumber
\end{eqnarray}

This defines a metric for each value of $s$ and $s^{*}$, such that the $%
\omega $ 's form a null tetrad. We wish to know how ($\alpha ,a,b,c)$ should
be specified for the $(s,s^{*})$-dependent metrics to be ``almost''
conformally equivalent for all ($s,s^{*}$). The metrics are said to be ``%
\textit{almost}'' conformally equivalent if 
\begin{eqnarray}
Dg &=&U_{ij}[M[S,S^{*}]]\omega ^{i}\otimes \omega ^{j}+\Lambda [S,S^{*}]g,
\label{minimal2} \\
D^{*}g &=&U_{ij}^{*}[M^{*}[S,S^{*}]]\omega ^{i}\otimes \omega ^{j}+\Lambda
^{*}[S,S^{*}]g,  \nonumber
\end{eqnarray}
where, (1); $\Lambda $ and $\Lambda ^{*}$ are explicit functions of ($%
S,S^{*})$, (2); $M[S,S^{*}]$ and $M^{*}[S,S^{*}]$ are specific non-linear
functions of ($S,S^{*})$ and their derivatives, [the ``metricity
expressions'' or generalized Wunschmann conditions], (3); $U_{ij}[M]$ are
functions of $M,$ $DM$ and $D^{2}M$ that all \textit{vanish} when $M=0.$ $S$
and $S^{*}$ are still arbitrary functions of $(Z,DZ,D^{*}Z,DD^{*}Z,s,s^{*}).$
The $(s,s^{*})$-dependent metrics are conformally equivalent when $S$ and $%
S^{*}$ are such that $M[S,S^{*}]\ $vanishes$.$ For arbitrary $S,$ however,
the metrics are ``almost'' conformally related. We refer to (\ref{minimal2})
as the ``minimal dependence conditions''.

In this section we display the values of ($\alpha ,a,b,c)$ in terms of $S,$
that satisfy the Eqs.(\ref{minimal2}). We could do this by using, \textit{in
principle} a ``simplicity'' argument (i.e., by trying to do the simplest
thing possible), which would consist of setting to zero certain components
of $Dg-\lambda g,$ (for some $\lambda );$ \textit{namely} those that allow
us to solve for ($\alpha ,a,b,c)$ \textit{algebraically} in terms of ($%
S,S^{*})$ and their derivatives. The remaining components of $Dg-\lambda g$ 
\textit{were then} to be then shown to be of the form $U_{ij}[M[S,S^{*}]]$
depending on a single function $M[S,S^{*}]$ which vanish when $M[S,S^{*}]$
vanishes. In this manner, the unknown functions ($\alpha ,a,b,c)$ \textit{%
were} \textit{to be uniquely} \textit{determined}. In fact we did not do
this. We did start this calculation and did, in this manner, determine,%
\textbf{\ }($\alpha ,a,b),$ (see below) but soon the complexity of the
algebraic expressions and manipulations became unmanageable and we could not
determine\textbf{\ }$c$ and $M[S,S^{*}]$ directly. There however was an
alternative approach (See Sec.IV) that did allow us to finish the task.

In the following we will first state the main results (partially obtained by
both methods) and then outline the ``simplicity'' argument. The results will
then be discussed in subsection B. Finally, in Sec.IV, the alternative
method will be described in detail. The equivalence of both methods is then
shown.

The main results are the following determination of the unknown functions%
\textbf{\ }($\alpha ,a,b,c);$

\begin{eqnarray}
b &=&\frac{1}{S_{R}^{*}}(\sqrt{1-S_{R}^{*}S_{R}}-1),\qquad b^{*}=\frac{1}{%
S_{R}}(\sqrt{1-S_{R}^{*}S_{R}}-1),  \label{bb*} \\
\alpha ^{2} &=&\frac{(\sqrt{1-S_{R}^{*}S_{R}}+1)}{2(1-S_{R}^{*}S_{R})}=\frac{%
(1+bb^{*})}{(1-bb^{*})^{2}}  \label{alpha} \\
a &=&(1-S_{R}S_{R}^{*})^{-1}(1-\frac{1}{4}S_{R}S_{R}^{*})^{-1}{\large \{}%
\frac{1}{2}[S_{W^{*}}^{*}+S_{W}^{*}S_{R}-T_{R}^{*}](1+\frac{1}{2}%
S_{R}^{*}S_{R})  \nonumber \\
&&-\frac{3}{4}S_{R}^{*}[S_{W}+S_{W^{*}}S_{R}^{*}-T_{R}]{\large \}}  \label{a}
\\
c &=&-\frac{1}{2}G-(a-a^{*}b^{*})(a^{*}-ab)(1+bb^{*})^{-1}.  \label{G}
\end{eqnarray}
where $T\equiv D^{*}S,$ $U\equiv D^{*}T=D^{2*}S\equiv D^{2}S^{*},$ the
subscripts on the $S,T,U$ refer to partial derivatives. $G$ is defined by

\begin{eqnarray}
G(1+\frac{1}{2}S_{R}S_{R}^{*})
&=&T_{W}+T_{W^{*}}S_{R}^{*}+T_{W^{*}}^{*}+T_{W}^{*}S_{R}-\frac{1}{2}U_{R}
\label{c} \\
&&+\frac{1}{2}%
(S_{W}^{*}S_{W}S_{R}+S_{W}S_{W^{*}}^{*}+S_{W^{*}}^{*}S_{W^{*}}S_{R}^{*}+S_{W^{*}}S_{W}^{*}
\nonumber \\
&&-S_{R}^{*}S_{Z}-S_{R}S_{Z}^{*})-\frac{1}{2}%
(S_{W}S_{R}^{*}+S_{R}S_{W}^{*}+2T_{R}^{*})\frac{g^{1+}}{g^{01}}  \nonumber \\
&&-\frac{1}{2}(S_{R}S_{W^{*}}^{*}+S_{W^{*}}S_{R}^{*}+2T_{R})\frac{g^{1-}}{%
g^{01}}.  \nonumber
\end{eqnarray}
with 
\begin{eqnarray}
\frac{g^{1+}}{g^{01}}(1-\frac{1}{4}S_{R}S_{R}^{*}) &=&-\frac{1}{2}[
T_{R}-S_{W}-S_{W^{*}}S_{R}^{*}]+\frac{1}{4}%
S_{R}[T_{R}^{*}-S_{W^{*}}^{*}-S_{W}^{*}S_{R}],  \nonumber \\
\frac{g^{1-}}{g^{01}}(1-\frac{1}{4}S_{R}S_{R}^{*}) &=&-\frac{1}{2}[%
T_{R}^{*}-S_{W^{*}}^{*}-S_{W}^{*}S_{R}]+\frac{1}{4}%
S_{R}^{*}[T_{R}-S_{W}-S_{W^{*}}S_{R}^{*}].  \nonumber
\end{eqnarray}

The metricity expression is given by

\begin{equation}
M[S,S^{*}]=\frac{-2}{(1-bb^{*})}\{Db+S_{W^{*}}-bS_{W}-(S_{R}+b)(a^{*}-ab)\}
\label{basicmetricity}
\end{equation}
with $b$ and $b^{*}$ given by Eq.(\ref{bb*}).

Explicitly the simplicity argument is carried out as follows: We begin by
constructing 
\begin{equation}
Dg=\eta _{ij}D\omega ^{i}\otimes \omega ^{j}+\eta _{ij}\omega ^{i}\otimes
D\omega ^{j}.  \label{Dg1}
\end{equation}
Working out, (see appendix), via Eqs.(\ref{one-forms}) and (\ref{betas}),
all the

\begin{equation}
D\omega ^{i}=A_{j}^{i}\omega ^{j},  \label{Dw}
\end{equation}
with $A_{j}^{i}$ explicit functions of the derivatives of $S$ and $S^{*}$
and the unknown functions ($\alpha ,a,b,c),$ we obtain

\begin{equation}
Dg=[\eta _{kj}A_{i}^{k}+\eta _{ik}A_{j}^{k}]\omega ^{i}\otimes \omega
^{j}\equiv G_{ij}\omega ^{i}\otimes \omega ^{j},  \label{Dg2}
\end{equation}
with symmetric $G_{ij}.$ The $G_{ij}$ are thus also explicitly known
functions (see appendix) of $S$ and $S^{*}$ and their derivatives and the ($%
\alpha ,a,b,c).$ ($Dg$ should be thought of as the Lie derivative of the
metric along the vector field defined by (\ref{totalD}).)

We now rewrite $Dg$ by adding and subtracting a term $2G_{01}\omega
^{(+}\otimes \omega ^{-)}$, obtaining

\begin{eqnarray}
Dg &=&G_{01}g+G_{11}\omega ^{1}\otimes \omega ^{1}+2G_{1+}\omega
^{(1}\otimes \omega ^{+)}+2G_{1-}\omega ^{(1}\otimes \omega ^{-)}
\label{Dg3} \\
&&+G_{--}\omega ^{-}\otimes \omega ^{-}+G_{++}\omega ^{+}\otimes \omega
^{+}+2(G_{01}+G_{+-})\omega ^{(+}\otimes \omega ^{-)}  \nonumber \\
&&+2G_{0+}\omega ^{(0}\otimes \omega ^{+)}+2G_{0-}\omega ^{(0}\otimes \omega
^{-)}+G_{00}\omega ^{0}\otimes \omega ^{0}  \nonumber
\end{eqnarray}

A simple inspection of the explicit expressions for $G_{ij}=[\eta
_{kj}A_{i}^{k}+\eta _{ik}A_{j}^{k}]$ and its conjugate (See the appendix)
reveals that, contrary to the 2+1 case of Section II, determining which
combinations of them should vanish is not at all, in this case, obvious .
However, guided by the procedure that leads to the Null Surface
reformulation of GR, as explained in Section IV, we take the following steps.

1. We observe that $G_{11}\equiv 0$.

2. Setting 
\begin{equation}
G_{1+}[S,b,\alpha]=G_{1-}[S,b,\alpha]=0  \label{G1pm0}
\end{equation}
\noindent determines $b(S)$ and $\alpha(S)$ \textit{algebraically}, as given
in Eqs.(\ref{bb*}) and (\ref{alpha}).

3. If $b(S)$ is given as in Eqs.(\ref{bb*}), then setting

\begin{equation}
G_{++}[S,b(S),\alpha(S),a]=b^{*2}(S)G_{--}[S,b(S),\alpha(S),a]  \label{G++0}
\end{equation}
allows us to determine the functions $a(S)$ \textit{algebraically} as in
Eq.~(\ref{a}).

4. We then have, (\ref{G++0}), with (\ref{bb*}) and (\ref{alpha}), that 
\begin{eqnarray*}
&&G_{01}[S,b(S),\alpha (S),a(S)], \\
&&G_{-+}[S,b(S),\alpha (S),a(S)], \\
&&G_{--}[S,b(S),\alpha (S),a(S)]
\end{eqnarray*}
are now explicit functions of $S$ and moreover they satisfy 
\begin{eqnarray*}
&&G_{01}[S,b(S),\alpha (S),a(S)]+G_{-+}[S,b(S),\alpha (S),a(S)] \\
&=&b^{*}(S)G_{--}[S,b(S),\alpha (S),a(S)]
\end{eqnarray*}
\textit{as an identity\/}.

Up to this point, namely, imposing (\ref{G1pm0}) and (\ref{G++0}), we have
that $Dg$ is reduced to

\begin{eqnarray}
Dg &=&G_{01}g +G_{--} {\large (}\omega ^{-}\otimes \omega ^{-} +b{^{*}}
^{2}(S)\omega^{+}\otimes \omega ^{+} +2b^{*}(S)(\omega ^{(+}\otimes \omega
^{-)} {\large )}  \nonumber \\
&&+2G_{0+}\omega^{(0}\otimes\omega^{+)} +2G_{0-}\omega ^{(0}\otimes
\omega^{-)} +G_{00} \,\omega ^{0}\otimes \omega ^{0}
\end{eqnarray}

\noindent where

\begin{eqnarray}
G_{01} &=&G_{01}[S,a(S)],  \nonumber \\
G_{--} &=&G_{--}[S,b(S),\alpha (S),a(S)],  \nonumber \\
G_{0+} &=&G_{0+}[S,b(S),\alpha (S),a(S),c],  \label{g's} \\
G_{0-} &=&G_{0-}[S,b(S),\alpha (S),a(S),c],  \nonumber \\
G_{00} &=&G_{00}[S,b(S),\alpha (S),a(S),c,Dc].  \nonumber
\end{eqnarray}

5. We now need to extract, from our minimal dependence condition (\ref
{minimal2}), a linear combination of the remaining components of $Dg-G_{01}g$
which, when vanishing, will allow us to obtain $c$ \textit{algebraically}
and \textit{simultaneously force} $G_{0+},G_{0-}$ and $G_{00}$ to vanish
when $G_{--}[S,b(S),\alpha (S),a(S)]=0$. The linear combination of $G$ 's
that determines $c$ in this way is

\begin{equation}
(2-bS_{R}^{*})G_{0+}+(S_{R}^{*}-2b^{*})G_{0-}+\alpha
^{-1}(1-bb^{*})(aS_{R}^{*}-S_{W}^{*})G_{--}=0  \label{G0+0}
\end{equation}

\noindent with $G_{0+},G_{0-}$ and $G_{--}$ given by (\ref{g's}).

This result becomes extremely difficult to see by simple inspection of the
equations. Instead, one must turn to the methods of Sec.IV. From this
analysis one could see that $c$ was given by (\ref{G})\ and that\textbf{\ }$
G_{0+},$ $G_{0-},$ $G_{00}$ all vanish when\textbf{\ }$G_{--}=0.$

Based on this, we promote $G_{--}[S,b(S),\alpha (S),a(S)]$ to the \textit{%
metricity condition} and make the following identifications:

\begin{equation}
G_{01}(S,a(S))\equiv \Lambda (S)
\end{equation}
and

\begin{equation}
G_{--}[S,b(S),\alpha (S),a(S)]\equiv M(S).
\end{equation}

{}From this we then have the form of $G_{ij}$ given in our minimal
dependence equation (\ref{minimal2}). When $M(S)=0$ we have that

\[
Dg=\Lambda [S]g. 
\]

\subsection{Six Dimensional Metrics}

We now extend the metric $g$ to a six-dimensional metric by

\begin{equation}
g^{(6)}(x^{a},s,s^{*})=g-\omega \otimes \omega ^{*}=g-CC^{*}ds\otimes ds^{*}
\end{equation}
so that 
\begin{eqnarray}
Dg^{(6)} &=&Dg-(C^{*}DC+CdC^{*})ds\otimes ds^{*}, \\
&=&U_{ij}[M[S,S^{*}]]\omega ^{i}\otimes \omega ^{j}+\Lambda
[S,S^{*}]g-(C^{*}DC+CDC^{*})ds\otimes ds^{*},  \nonumber
\end{eqnarray}
If the unknown $C$ is chosen so that $DC=\frac{1}{2}\Lambda C$ , $DC^{*}=%
\frac{1}{2}\Lambda C^{*}$ then 
\begin{equation}
Dg^{(6)}=U_{ij}[M[S,S^{*}]]\omega ^{i}\otimes \omega ^{j}+\Lambda g^{(6)}. 
\nonumber
\end{equation}
This leads to the

\begin{proposition}
If the class of differential equations is restricted to those $S$ that
satisfy the conditions 
\begin{equation}
M[S,S^{*}]=0,\text{ }M^{*}[S,S^{*}]=0,  \label{metricity}
\end{equation}
then there exists on the six-space, a pair of conformal Killing fields $\xi
=d/ds$ and $\xi ^{*}=d/ds^{*}.$ It is obvious that conformal factors $\Omega 
$ can easily be found for the six metric so that the conformal Killing
fields become Killing fields and the six-space can be foliated with
four-dimensional subspaces, ($s$ and $s^{*}$ constant), with the induced
four metrics all isometric. These four metric then map down to a unique
conformal class of Lorentzian metric on the four-space of the $x^{a}.$
\end{proposition}

\section{Relationship with the Null Surface Formulation of GR}

It is not hard to see that all Lorentzian 4-metrics (locally) are included
in this construction. This follows from its equivalence to the Null Surface
reformulation of General Relativity. In that work\cite{S1,S2,S3,S4}, one
begins with a four-manifold $\frak{M}$ with an unknown - but to be
determined - \textit{conformal} Lorentzian metric and ask that there be a
two-parameter, $(s,s^{*}),$ family of (local) null surface foliations of $%
\frak{M}$ of sufficient generality, whose level surfaces are given by

\[
u=Z(x^{a},s,s^{*}). 
\]
This requires that the \textit{unknown conformal metric} satisfies 
\begin{equation}
g^{ab}\partial _{a}Z\partial _{b}Z=0  \label{NULL}
\end{equation}
\textit{for all} ($s,s^{*}).$ The arbitrary conformal factor can depend on $%
(s,s^{*}).$ By repeated $(s,s^{*})$ derivatives of Eq.(\ref{NULL}),
(explicitly, the following eight derivatives) 
\[
D,\text{ }D^{*},\text{ }D^{2},\text{ }D^{*2},\text{ }DD^{*},\text{ }%
D^{*}D^{2},\text{ }DD^{*2},\text{ }D^{2}D^{*2}, 
\]
which with Eq.(\ref{NULL}), yields nine relations so that the unknown
conformal metric, $g^{ab},$ can be given completely in terms of a function 
\[
S(Z,DZ,D^{*}Z,DD^{*}Z,s,s^{*}) 
\]
that is defined by

\[
D^{2}Z=S,\text{ }D^{*2}Z=S^{*} 
\]
and we are back to our starting point. These metrics satisfy our minimal
dependence condition, Eq.(\ref{minimal2}), for ($s,s^{*})$ dependence. If we
continue and take the derivatives $D^{3}$ and $D^{*3}$ of Eq.(\ref{NULL}),
we finally obtain the \textit{metricity conditions} $M[S,S^{*}]=0,$ $%
M^{*}[S,S^{*}]=0.$

Since we began with an arbitrary Lorentzian space we see that for any such
space there exists an $S$ satisfying the metricity conditions that yields,
via Eqs.(\ref{pair2}), that metric up to conformal factor.

It is now clear that the vacuum Einstein equations, with a specific choice
of conformal factor, can be obtained by a further restriction on the class
of functions ($S,S^{*})$.

The procedure just outlined can be carried out explicitly, with considerable
calculational effort, as follows:

Start with a function $Z=Z(x^{a},s,s^{*})$ and its $D$, $D^{*}$ and $DD^{*}$
derivatives as ``primary'' functions, i.e., $\theta ^{i}=(\theta ^{0},\theta
^{+},\theta ^{-},\theta ^{1})\equiv (Z,W,W^{*},R);$

\begin{eqnarray}
Z &=&Z(x^{a},s,s^{*}),  \label{P} \\
W &=&DZ(x^{a},s,s^{*}),  \nonumber \\
W^{*} &=&D^{*}Z(x^{a},s,s^{*}),  \nonumber \\
R &=&D^{*}DZ(x^{a},s,s^{*})=D^{*}W=DW^{*}.  \nonumber
\end{eqnarray}

It is assumed that these four relations can be inverted as $%
x^{a}=X^{a}(\theta ^{i},s,s^{*}).$

The set of ``secondary'' functions 
\begin{eqnarray}
S &\equiv &D^{2}Z,\ \quad \quad S^{*}\equiv D^{*2}Z,  \label{S} \\
T &\equiv &DR=D^{*}S=D^{*}D^{2}Z,\ \quad T^{*}\equiv D^{*}R=DS^{*}=DD^{*2}Z,
\nonumber \\
U &\equiv &D^{2}S^{*}=D^{*2}S=D^{*}T=DT^{*}=D^{2}D^{*2}Z,  \nonumber
\end{eqnarray}
which can all be thought of as functions of $(Z,W,W^{*},R,s,s^{*})$ where
the $x^{a}$ have been eliminated via the inversion of Eq.(\ref{P}). The
exterior derivatives (the space-time gradients, holding ($s,s^{*})$ constant
) of the ``primary'' functions are 
\begin{equation}
d\theta ^{i}=\partial _{a}\theta ^{i}dx^{a}  \label{dtheta}
\end{equation}
and for the ``secondary'' functions $S,T$ and $U,$ they are given by

\begin{eqnarray}
dS &=&S_{Z}dZ+S_{W}dW+S_{W^{*}}dW^{*}+S_{R}dR=S_{\theta ^{i}}d\theta ^{i},
\label{dS} \\
dT &=&T_{Z}dZ+T_{W}dW+T_{W^{*}}dW^{*}+T_{R}dR=T_{\theta ^{i}}d\theta ^{i}, 
\nonumber \\
dU &=&U_{Z}dZ+U_{W}dW+U_{W^{*}}dW^{*}+U_{R}dR=U_{\theta ^{i}}d\theta ^{i}. 
\nonumber
\end{eqnarray}

If we write the unknown, but to be determined, inverse of the space-time
metric $g^{ab}(x^{a},s,s^{*})=\widehat{g}^{ab}(x^{a})\omega ^{2}(s,s^{*}),$
(i.e.,where the $(s,s^{*})$ behavior appears only in the conformal factor)
as 
\[
g^{I}\equiv g^{ab}\partial _{a}\otimes \partial _{b} 
\]
then we can define the metric components in the gradient basis, $\theta
_{a}^{i}\equiv \partial _{a}\theta ^{i}$, by

\begin{equation}
g^{ij}(x^{a},s,s^{*})\equiv g^{I}(d\theta ^{i},d\theta ^{j})=g^{ab}\theta
_{a}^{i}\theta _{b}^{j}.  \label{ij}
\end{equation}

Note the very important point that since $g^{ab}=\widehat{g}%
^{ab}(x^{a})\omega ^{2}(s,s^{*})$ we have 
\begin{equation}
Dg^{I}\equiv Dg^{ab}\partial _{a}\otimes \partial _{b}=2\omega ^{-1}D\omega
g^{I}\equiv \lambda g^{I}  \label{DgI}
\end{equation}
We, however, will not be using Eq.(\ref{DgI}) fully until the end of the
calculation. More explicitly, we will only be using different specific
components of Eq.(\ref{DgI}), i.e., specific components of 
\begin{equation}
Dg^{\text{I}}(d\theta ^{i},d\theta ^{j})\equiv Dg^{ab}\theta _{a}^{i}\theta
_{\,b}^{j}=\lambda g^{ij}  \label{DgIij}
\end{equation}
along the way and only at the end, with the metricity condition, will the
full Eq.(\ref{DgI}) be used. Until this last condition is imposed the
conditions on $Dg^{I}$ are precisely our ``minimal dependence conditions''.

Starting with the condition that the level surfaces , $Z(x^{a},s,s^{*})=$ $%
constant,$ (for each value of $(s,s^{*})$) are null surfaces of the metric,
we have that 
\begin{equation}
g^{00}=g^{\text{I}}(dZ,dZ)=g^{ab}Z,_{a}Z,_{b}=0.  \label{00}
\end{equation}
By applying $D$ and $D^{*}$ to Eq.(\ref{00}), we have

\begin{eqnarray}
Dg^{ab}Z,_aZ,_b +2g^{ab}W,_aZ,_b =0, \\
D^*\!g^{ab}Z,_aZ,_b +2g^{ab}W^*\!\!\!,_aZ,_b =0.
\end{eqnarray}

\noindent Thus from Eq.(\ref{00}, using one component of Eq.(\ref{DgIij})

\begin{equation}
Dg^{ab}Z,_{a}Z,_{b}=\lambda g^{ab}Z,_{a}Z,_{b}=0,  \label{dg00}
\end{equation}

\noindent we have 
\begin{eqnarray}
g^{0+} &=&g^{I}(dDZ,dZ)=g^{I}(dW,dZ)=0,  \label{0+} \\
g^{0-} &=&g^{I}(dD^{*}Z,dZ)=g^{I}(dW^{*},dZ)=0.  \label{0-}
\end{eqnarray}

Next. applying $D$ to Eq.(\ref{0-}), yields

\begin{equation}
Dg^{ab}W^*\!\!\!,_aZ,_b +g^{ab}\left(R,_aZ,_b + W^*\!\!\!,_aW,_b\right) =0.
\end{equation}

\noindent Thus from Eq.(\ref{0-}), again with one component of Eq.(\ref
{DgIij})

\begin{equation}
Dg^{ab}W^{*}\!\!\!,_{a}Z,_{b}=\lambda g^{ab}W^{*}\!\!\!,_{a}Z,_{b}=0,
\label{dg0-}
\end{equation}

\noindent we have

\begin{equation}
g^{\text{I}}(dR,dZ)+g^{\text{I}}(dW,dW^{*})=0\quad \Leftrightarrow \quad
g^{-+}+g^{01}=0.  \label{01}
\end{equation}

If $D$ and $D^{*}$ are applied respectively to Eqs.(\ref{0+}) and (\ref{0-}
), we have 
\begin{eqnarray}
Dg^{ab}W,_{a}Z,_{b}+g^{ab}(S,_{a}Z,_{b}+W,_{a}W,_{b}) &=&0, \\
D^{*}\!g^{ab}W^{*}\!\!\!,_{a}Z,_{b}+g^{ab}(S^{*}\!\!\!,_{a}Z,_{b}+W^{*}\!\!%
\!,_{a}W^{*}\!\!\!,_{b}) &=&0.
\end{eqnarray}

\noindent Therefore, from Eq.(\ref{0+}) again using one component of Eq.(\ref
{DgIij}),

\begin{equation}
Dg^{ab}W,_{a}Z,_{b}=\lambda g^{ab}W,_{a}Z,_{b}=0,  \label{dg0+}
\end{equation}

\noindent (which implies its complex conjugate as well) then, using (\ref{dS}%
),

\begin{eqnarray}
g^{I}(dS,dZ)+g^{I}(dW,dW) &=&0\quad \Leftrightarrow g^{++}=-S_{R}g^{01}\quad
\label{++} \\
g^{I}(dS^{*},dZ)+g^{I}(dW^{*},dW^{*}) &=&0\quad \Leftrightarrow
g^{--}=-S_{R}^{*}g^{01}  \label{--}
\end{eqnarray}

Continuing this process, i.e., applying $D$ and $D^{*}$ to Eq.(\ref{01}),
yields

\begin{eqnarray}
Dg^{ab}W,_aW^*\!\!\!,_b +Dg^{ab}Z,_aR,_b +g^{ab}(S,_aW^*\!\!\!,_b +2W,_aR,_b
+Z,_aT,_b) =0, \\
D^*\!g^{ab}W,_aW^*\!\!\!,_b +D^*\!g^{ab}Z,_aR,_b +g^{ab}(W,_aS^*\!\!\!,_b
+2W^*\!\!\!,_aR,_b +Z,_aT^*\!\!\!,_b) =0.
\end{eqnarray}

\noindent Therefore, from Eq.(\ref{01}) using two components of Eq.(\ref
{DgIij}), we have

\begin{equation}
Dg^{ab}W,_{a}W^{*}\!\!\!,_{b}+Dg^{ab}Z,_{a}R,_{b}=\lambda
g^{ab}(W,_{a}W^{*}\!\!\!,_{b}+Z,_{a}R,_{b})=0,  \label{dg+-}
\end{equation}

\noindent (and its complex conjugate) leads, respectively, to

\begin{eqnarray}
g^{I}(dT,dZ)+g^{I}(dS,dW^{*})+g^{I}(dR,dW)+g^{I}(dW,dR) &=&0,  \label{g1+} \\
g^{I}(dT^{*},dZ)+g^{I}(dR,dW^{*})+g^{I}(dR,dW^{*})+g^{I}(dW,dS^{*}) &=&0.
\label{g1-}
\end{eqnarray}

\noindent Using Eqs.(\ref{dS}), (\ref{00}),(\ref{0+}),(\ref{0-}) and (\ref
{++}), in (\ref{g1+}) and (\ref{g1-}) we obtain, respectively,

\begin{eqnarray}
2g^{+1} &=&-T_{R}g^{01}+S_{W}g^{01}+S_{W^{*}}S_{R}^{*}g^{01}-S_{R}g^{-1},
\label{1+} \\
2g^{-1}
&=&-T_{R}^{*}g^{01}+S_{W^{*}}^{*}g^{01}+S_{W}^{*}S_{R}g^{01}-S_{R}^{*}g^{+1},
\nonumber
\end{eqnarray}
which are easily solved for $g^{+1}$ and $g^{-1}$:

\begin{eqnarray}
g^{+1}(1-\frac{1}{4}S_{R}S_{R}^{*}) &=&-\frac{1}{2}[
T_{R}-S_{W}-S_{W^{*}}S_{R}^{*}]g^{01}  \label{1-final} \\
&&+\frac{1}{4}S_{R}[T_{R}^{*}-S_{W^{*}}^{*}-S_{W}^{*}S_{R}]g^{01},  \nonumber
\\
g^{-1}(1-\frac{1}{4}S_{R}S_{R}^{*}) &=&-\frac{1}{2}[%
T_{R}^{*}-S_{W^{*}}^{*}-S_{W}^{*}S_{R}]g^{01}  \label{1+final} \\
&&+\frac{1}{4}S_{R}^{*}[T_{R}-S_{W}-S_{W^{*}}S_{R}^{*}]g^{01}.  \nonumber
\end{eqnarray}

\noindent Notice that Eqs.~(\ref{g1+}) and (\ref{g1-}) are obtained just as
well by taking $D^{*}$ of Eq.~(\ref{++}) and $D$ of (\ref{--}) using Eq.(\ref
{--}) and two components of Eq.(\ref{DgIij}), we have

\begin{equation}
Dg^{ab}W^*\!\!\!,_aW^*\!\!\!,_b +Dg^{ab}S,_aZ,_b =0  \label{dg--}
\end{equation}

\noindent (with its complex conjugate).

Finally, by applying $D^{*}$ to Eq.(\ref{g1+}), (or $D$ to Eq.(\ref{g1-})),
we obtain the last metric component $g^{11};$

\begin{eqnarray}
-2(1+\frac{1}{2}S_{R}S_{R}^{*})g^{11}
&=&(2T_{W}^{*}+S_{W}S_{W}^{*})g^{++}+(2T_{W^{*}}+S_{W^{*}}S_{W^{*}}^{*})g^{--}
\label{11*} \\
&&+g^{01}(U_{R}+S_{Z}S_{R}^{*}+S_{R}S_{Z}^{*}  \nonumber \\
&&-S_{W^{*}}S_{W}^{*}-S_{W}S_{W^{*}}^{*}-2T_{W^{*}}^{*}-2T_{W})  \nonumber \\
&&+g^{-1}(S_{R}S_{W^{*}}^{*}+S_{W^{*}}S_{R}^{*}+2T_{R})  \nonumber \\
&&+g^{+1}(S_{R}S_{W}^{*}+S_{W}S_{R}^{*}+2T_{R}^{*})  \nonumber
\end{eqnarray}

\noindent using the equation

\begin{equation}
Dg^{ab}(T^{*}\!\!\!,_{a}Z,_{b}+S^{*}\!\!\!,_{a}W,_{b}+2R,_{a}W^{*}\!\!%
\!,_{b})=0.  \label{dg1-}
\end{equation}
that arises from similar considerations, from Eq.(\ref{DgIij}), as before.

We emphasize that at this point we have not yet used the full set of
components of Eq.(\ref{DgIij}) and consequently we do not yet have a single
conformal metric by this construction - but instead we have an $(s,s^{*})$
dependent family of conformal metrics. In other words we see that modulo an
overall (conformal) factor, namely $g^{01}=\omega ^{2}(s,s^{*}),$ a two
parameter family of metrics, $g^{ij}(x^{a},s,s^{*}),$ has been obtained by
requiring that a series of $D$ and $D^{*}$ derivatives, applied to the null
surface condition, Eq.(\ref{00}) remains zero for all $(s,s^{*})$ [this is
the meaning of imposing Eqs.~(\ref{dg00}), (\ref{dg0-}),(\ref{dg0+}),(\ref
{dg+-}),(\ref{dg--}) and (\ref{dg1-})]. \textit{All the components, in the
gradient basis,} $\theta _{a}^{i}$, \textit{have been expressed in terms of
derivatives of} $S$ \textit{and} $S^{*}$. This $(s,s^{*})-$dependent metric
satisfies the minimal dependence condition, Eq.(\ref{minimal2}), and is, in
fact, identical (up to a conformal rescaling) to the metric of Eq.(\ref
{4metric}), but is expressed in a gradient basis, rather than in the null
tetrad basis. By using one further component of Eq.(\ref{DgIij}), namely
applying $D$ to Eq.(\ref{++}) and using

\begin{equation}
Dg^{ab}(W,_{a}W,_{b}+S,_{a}Z,_{b})=0,
\end{equation}
we obtain the metricity condition the only condition of the functions $S$
and $S^{*}$. (Given below.) In this case the $(s,s^{*})-$ dependent family
of metric are all conformal to each other.

We notice that all the components of the metric, $g^{ij},$ are determined up
to a single overall undetermined factor, namely $g^{01};$ i.e., $%
g^{ij}=g^{01}h^{ij}[S,S^{*}].$ To make the explicit comparison between this
conformal metric and the metric, (\ref{4metric}), we chose the special
conformal gauge

\[
g^{01}=1. 
\]
If we take the null tetrad system, $\omega ^{i}\!_{a}$ as a linear
combination of the gradient basis $\theta ^{j}\!,_{b}$ then re-expressing
the metric in the null tetrad system, $\omega ^{i}\!_{a},$ allows us to read
off the coefficients, $(\alpha ,a,b,c)$, of Eq.(\ref{one-forms}). Explicitly,

\begin{equation}
g_{ab}=g_{ij}\theta ^{i}\!,_{a}\theta ^{j}\!,_{b}=\eta _{ij}\omega
^{i}\!_{a}\omega ^{j}\!_{b}=\eta _{ij}K_{k}^{i}K_{l}^{j}\theta
^{k}\!,_{a}\theta ^{l}\!,_{b},
\end{equation}

\noindent or

\begin{equation}
\eta_{kl} K^k_i K^l_j = g_{ij},  \label{nullcoeff}
\end{equation}

\noindent where the coefficients $K_{j}^{i}$ are given by $\omega
_{a}^{i}\equiv K_{j}^{i}\theta ^{j}\!,_{a}$ and are shown explicitly in Eqs.
(\ref{one-forms}). In particular, $\alpha =K_{+}^{+},$ $a=K_{+}^{1},$ $%
b=K_{-}^{+}/K_{+}^{+}$ and $c=K_{0}^{1}$. From (\ref{nullcoeff}), then using
the special conformal frame, ($g^{01}=1\Rightarrow g^{+-}=-1)$, we obtain

\begin{eqnarray*}
b &=&\frac{-1}{g^{--}}(\sqrt{J}-1), \\
b^{*} &=&\frac{-1}{g^{++}}(\sqrt{J}-1), \\
\alpha ^{2} &=&\frac{g^{++}g^{--}}{2J(\sqrt{J}-1)} \\
a &=&\frac{g^{1+}g^{--}+g^{1-}}{J}, \\
a^{*} &=&\frac{g^{1-}g^{++}+g^{1+}}{J}, \\
c &=&-\frac{1}{2}g^{11}+\frac{[g^{--}g^{1+}+g^{1-}+g^{1-}\sqrt{J}%
][g^{++}g^{1-}+g^{1+}+g^{1+}\sqrt{J}]}{2J(\sqrt{J}-1)},
\end{eqnarray*}
with 
\[
J=(g^{-+})^{2}-g^{++}g^{--}=1-S_{R}S_{R}^{*}. 
\]

\noindent Since the components of the metric are functions of $S$ as
obtained above, the parameters $(\alpha,a,b,c)$ are expressed in terms of $S$
, as desired.

We can now see how this procedure justifies Eqs.~(\ref{G1pm0}), (\ref{G++0})
and (\ref{G0+0}) used in Section III. Straightforward algebra shows that

\begin{equation}
Dg^{ab}\theta^i_a\theta^j_b = -\eta^{kl}G_{km}\eta^{mn}(K^{-1})^i_l
(K^{-1})^j_n  \label{basic}
\end{equation}

Using (\ref{basic}) to translate Eqs.~(\ref{dg00}), (\ref{dg0-}), (\ref{dg0+}
), (\ref{dg+-}), (\ref{dg--}) and (\ref{dg1-}) in terms of $G_{ij}$, the
following propositions follow:

\begin{proposition}
The vanishing of $Dg^{ab}\theta _{a}^{0}\theta _{b}^{0}$ is equivalent to
the vanishing of $G_{11}$. Explicitly, 
\begin{equation}
Dg^{ab}\theta _{a}^{0}\theta _{b}^{0}=0\Leftrightarrow G_{11}=0.
\end{equation}
\end{proposition}

\begin{proposition}
The vanishing of $Dg^{ab}\theta _{a}^{-}\theta _{b}^{0}$ and $Dg^{ab}\theta
_{a}^{+}\theta _{b}^{0}$ (with $G_{11}=0)$ is equivalent to the vanishing of 
$G_{1+}$ and $G_{1-}$. Thus 
\begin{equation}
Dg^{ab}\theta _{a}^{-}\theta _{b}^{0}=0=Dg^{ab}\theta _{a}^{+}\theta
_{b}^{0}\Leftrightarrow G_{1+}=0=G_{1-}.
\end{equation}
\end{proposition}

\begin{proposition}
The vanishing of $Dg^{ab}(\theta _{a}^{+}\theta _{b}^{-}+\theta
_{a}^{0}\theta _{b}^{1})$ and $Dg^{ab}(\theta _{a}^{-}\theta
_{b}^{-}+S_{R}^{*}\theta _{a}^{0}\theta _{b}^{1})$ (with $%
G_{11}=G_{1+}=G_{1-}=0)$ is equivalent to the vanishing of $[G_{++}-{b^{*}}%
^{2}G_{--}]$ and $[G_{+-}+G_{01}-b^{*}G_{--}]$. 
\begin{eqnarray}
Dg^{ab}(\theta _{a}^{+}\theta _{b}^{-}+\theta _{a}^{0}\theta _{b}^{1})
&=&Dg^{ab}(\theta _{a}^{-}\theta _{b}^{-}+S_{R}^{*}\theta _{a}^{0}\theta
_{b}^{1})=0 \\
&\Leftrightarrow &[G_{++}-{b^{*}}^{2}G_{--}]=[G_{+-}+G_{01}-b^{*}G_{--}]=0.
\end{eqnarray}
\end{proposition}

This justifies our choices of vanishing combinations of $G_{ij}$ in order to
obtain $(\alpha,a,b,c)$ in terms of $S$ in Section III.

The metricity condition, from this point of view, is obtained simply by
applying $D$ to Eq.(\ref{++}), i.e. to 
\[
g^{I}(dS,dZ)+g^{I}(dW,dW)=0, 
\]

\noindent which, with the use of

\begin{equation}
Dg^{ab}(W,_{a}W,_{b}+S,_{a}Z,_{b})=0,
\end{equation}
which follows from the last component of (\ref{DgI}), leads to 
\[
g^{I}(dDS,dZ)+3g^{I}(dS,dW)=0, 
\]
or 
\[
M(S,S^{*})=\frac{1}{3}(DS)_{R}+S_{W}\frac{g^{++}}{g^{01}}+S_{W^{*}}\frac{%
g^{-+}}{g^{01}}+S_{R}\frac{g^{1+}}{g^{01}}=0. 
\]

\noindent Furthermore, with the use of (\ref{basic}) we can prove the
following

\begin{proposition}
The vanishing of $Dg^{ab}(\theta _{a}^{+}\theta _{b}^{+}+S_{R}\theta
_{a}^{0}\theta _{b}^{1})$ (with the earlier conditions) is equivalent to the
vanishing of $G_{--}$. Thus if $G_{11}=G_{1+}=G_{1-}=0$, $G_{++}={b^{*}}%
^{2}G_{--}$ and $G_{+-}+G_{01}=b^{*}G_{--}$, then 
\begin{equation}
Dg^{ab}(\theta _{a}^{+}\theta _{b}^{+}+S_{R}\theta _{a}^{0}\theta
_{b}^{1})=0\Leftrightarrow G_{--}=0.  \label{MM}
\end{equation}
Note that $G_{--}=0$ implies $G_{+-}+G_{01}=0.$
\end{proposition}

\noindent This allows us to promote $G_{--}$ to the status of metricity
condition if $(\alpha ,a,b,c)$ are given in terms of $S$, as argued in
Section~III.

When the metricity condition, Eq.(\ref{MM}), is imposed the expressions for
the remaining components of Eq.(\ref{DgIij}) are identically satisfied and 
\textbf{$G_{0-}$ }and\textbf{\ $G_{00}$ }vanish.

When $G_{--}\neq 0,$ the expression from Eq.(\ref{minimal2})

\[
U_{ij}[M[S,S^{*}]] 
\]
is a linear combination of $M(S,S^{*}),$ $D^{*}M(S,S^{*})$ and $%
D^{2*}M(S,S^{*})$ which thus all vanish when $M(S,S^{*})$ vanishes.

\section{Discussion}

In this work we have extended Cartan's beautiful construction of
differential geometric structures that are naturally associated with
ordinary differential equations, to a pair of overdetermined partial
differential equations. The resulting geometric structures form a rich set
of mathematical constructions that includes as a special case all Lorentzian
space-times $-$ and, as an obvious consequence, all solutions of Einstein's
theory of general relativity.

The study of the Einstein equations via this approach had already begun in
an earlier series of papers\cite{S1,S2,S3,S4} long before we connected it
with Cartan's view. This work on general relativity is continuing with
hopefully many applications, but the point of view towards it diverges from
that of the present paper. Here we feel that the issues are more closely
related to the study of equivalence classes of the starting equations under
some set or class of transformations.

An immediate question that we wish to investigate is the following; given
our set of equations

\[
D^{2}Z=S,\ \quad \quad D^{*2}Z=S^{*} 
\]
where the $(S,S^{*})$ satisfy the metricity conditions, the level surfaces
of the (local) solutions, $u=Z(x^{a},s,s^{*}),$ define a two-parameter
family of null surfaces (a complete integral) in the space-time of the
solution space, $x^{a},$ with the (conformal) metric, Eq.(\ref{4metric}). In
general, null surfaces develop caustics or wavefront singularities at which
point the function $Z(x^{a},s,s^{*})$ no longer satisfies the differential
equations - the local solutions break down. Nevertheless the space-time and
its metric could be completely smooth there. Other families of null surfaces
would exist that satisfied similar equations but with different $(S,S^{*}).$
We will consider the ``restricted equivalence'' problem to be the problem of
finding all pairs of functions, $(S,S^{*})$, that yield the same space-time
metric.\ As a matter of fact, this ``restricted problem'' is intimately
related with the equivalence problem under general contact transformations.
A paper is being prepared on this issue\cite{Niky}.

Another issue that we left untreated was how do the Cartan structure
equations, Eq.~(\ref{Structure}), behave in the case of the pair of PDE's.
It seems very likely that we will have the similar results to that obtained
from the third order ODE of Sec.II, where the metricity function plays the
role of a torsion tensor and we obtain a conformal connection rather than a
metric connection. But this remains to be analyzed - the calculations being
quite lengthy.

\section{Acknowledgments}

The authors thank the NSF for support under research grants \# PHY 92-05109,
PHY 97-22049 and PHY 98-03301. Carlos Kozameh thanks CONICET for support.

ETN is endebted to Peter Vassiliou, Niky Kamran and Robert Bryant for
enlightening discussions and a detailed proof of the four-dimensionality of
the solutions space for our pair of pdes.

We thank Paul Tod for both pointing out to us the connection of our earlier
work to the work of Cartan and Chern that got us started in this project,
for his teaching us Cartan's construction for 2nd order ODE's and for his
helpful comments on an early version of this manuscript.

\section{Appendix}

\subsection{Geometry of $\frac{d^{2}}{ds^{2}}$z = E(z,$\frac{dz}{ds}$,s)}

We begin with an arbitrary function of four variables 
\begin{equation}
\Phi (z,s,u,v)=0  \label{phi}
\end{equation}
assuming that it can be locally solved for any one of the variables.
Considering the two two-spaces of $(z,s)$ and $(u,v)$, we see that a point, $%
(z,s)$ in the first, corresponds to a specific curve in the second $-$ as
well as the converse. If we solve Eq.(\ref{phi}) for 
\begin{equation}
z=z(s,u,v)\equiv z(s,x^{A})  \label{sol}
\end{equation}
by differentiating with respect to $s,$ ($u,v$) can be eliminated from the
second derivative leaving

\[
z^{\prime \prime }=E(z,z^{\prime },s). 
\]

The same thing can be done with the variables ($u,v$) resulting in the
second order differential equation 
\begin{equation}
\frac{d^{2}u}{dv^{2}}=U(u,\frac{du}{dv},v).  \label{geod}
\end{equation}
for the curve in the ($u,v$) space.

Cartan then asks for the conditions on $E(z,z^{\prime },s)$ such that Eq.(%
\ref{geod}) is a geodesic for some (projective) symmetric connection - which
is determined by the form of $E$. He find that $E(z,z^{\prime },s)$ must
satisfy

\begin{equation}
\frac{d^{2}}{ds^{2}}E_{z^{\prime }z^{\prime }}-\frac{d}{ds}E_{zz^{\prime
}}-E_{z^{\prime }}\frac{d}{ds}E_{z^{\prime }z^{\prime
}}+2E_{zz}+E_{z^{\prime }}E_{zz^{\prime }}-2E_{z}E_{z^{\prime }z^{\prime }}=0
\label{M0}
\end{equation}
and the connection is given, in the gradient basis, $z_{A}\equiv \partial
_{A}z$ and $z_{A}^{\prime }\equiv \partial _{A}z^{\prime },$ by

\begin{eqnarray}
\nabla _{B}z_{A} &=&0,  \label{gradz} \\
\nabla _{B}z_{A}^{\prime } &=&2\alpha _{(A}^{\prime }z_{B)},  \label{gradz'}
\\
\alpha _{A}^{\prime } &=&\frac{1}{2}(E_{zz^{\prime }}-\frac{d}{ds}
E_{z^{\prime }z^{\prime }})z_{A}+E_{z^{\prime }z^{\prime }}z_{A}^{\prime },
\label{gradz''}
\end{eqnarray}
remembering the projective equivalence $\Gamma _{BC}^{A}\sim \Gamma
_{BC}^{A}+2\delta _{(B}^{A}\Upsilon _{C)}.$

This arises from the following argument: If the curve determined in Eq.(\ref
{sol}) by $(z,s)$ $=constant$ has a tangent vector $t^{A},$ then $t^{A}$ has
a vanishing product with the gradient of $z=z(s,x^{A});$ i.e., 
\begin{equation}
t^{A}z_{A}=0.  \label{tz}
\end{equation}

If $t^{A}$ is tangent to a geodesic then

\begin{equation}
t^{B}\nabla _{B}t^{A}=\beta t^{A}  \label{tgrad}
\end{equation}
and 
\[
t^{B}\nabla _{B}(z_{A}t^{A})=t^{A}t^{B}\nabla _{B}z_{A}+z_{A}t^{B}\nabla
_{B}(t^{A})=t^{A}t^{B}\nabla _{B}z_{A}=0. 
\]
Now since $z_{A}$ and $z_{A}^{\prime }$ form a basis set for the covectors,
we have that 
\[
\nabla _{B}z_{A}=az_{(B}z_{A)}+bz_{(B}^{\prime }z_{A)}+cz_{(B}^{\prime
}z_{A)}^{\prime } 
\]
but from Eqs.(\ref{tz}) and (\ref{tgrad}) we have $c=0$ and hence

\[
\nabla _{B}z_{A}=az_{(B}z_{A)}+bz_{(B}^{\prime }z_{A)}=\alpha _{(A}z_{B)} 
\]
and immediately 
\[
\nabla _{B}z_{A}^{\prime }=\alpha _{(A}^{\prime }z_{B)}+\alpha
_{(A}z_{B)}^{\prime }. 
\]
But via the projective equivalence they are the same as 
\begin{eqnarray}
\nabla _{B}z_{A} &=&0,  \label{covar} \\
\nabla _{B}z_{A}^{\prime } &=&\alpha _{(A}^{\prime }z_{B)}  \nonumber
\end{eqnarray}

By taking another $s$ (or prime) derivative 
\[
\nabla _{B}z_{A}^{\prime \prime }=\alpha _{(A}^{\prime \prime }z_{B)}+\alpha
_{(A}^{\prime }z_{B)}^{\prime } 
\]
and comparing it with 
\begin{eqnarray*}
\nabla _{B}z_{A}^{\prime \prime } &=&(E_{zz}\nabla z_{A}+E_{zz^{\prime
}}\nabla z_{A}^{\prime })z_{B} \\
&&+(E_{zz^{\prime }}\nabla _{A}z+E_{z^{\prime }z^{\prime }}\nabla
_{A}z^{\prime })z_{B}^{\prime }+E_{z^{\prime }}2\alpha _{(A}^{\prime }z_{B)}
\end{eqnarray*}
obtained from 
\[
z^{\prime \prime }=E(z,z^{\prime },s) 
\]
and

\[
z_{A}^{\prime \prime }=E_{z}z_{A}+E_{z^{\prime }}\,z_{A}^{\prime }, 
\]
we recover Eqs.(\ref{M0}) and (\ref{gradz''}).

We are indebted to Paul Tod for explaining this construction, due to Cartan,
to us.

\subsection{{\protect\large D}$\omega =A\omega $}

We have, via a lengthy calculation that the $A_{j}^{i},$ defined by $D\omega
^{i}=A_{j}^{i}\omega ^{j}$ or

\begin{eqnarray}
D\omega ^{0} &=&A_{0}^{0}\omega ^{0}+A_{+}^{0}\omega ^{+}+A_{-}^{0}\omega
^{-}+A_{1}^{0}\omega ^{1}  \label{notation} \\
D\omega ^{+} &=&A_{0}^{+}\omega ^{0}+A_{+}^{+}\omega ^{+}+A_{-}^{+}\omega
^{-}+A_{1}^{+}\omega ^{1},  \nonumber \\
D\omega ^{-} &=&A_{0}^{-}\omega ^{0}+A_{+}^{-}\omega ^{+}+A_{-}^{-}\omega
^{-}+A_{1}^{-}\omega ^{1},  \nonumber \\
D\omega ^{1} &=&A_{0}^{1}\omega ^{0}+A_{+}^{1}\omega ^{+}+A_{-}^{1}\omega
^{-}+A_{1}^{1}\omega ^{1},  \nonumber
\end{eqnarray}

are given by

\begin{eqnarray}
D\omega ^{0} &=&(1-bb^{*})^{-1}\alpha ^{-1}(\omega ^{+}-b\omega ^{-})
\label{startII} \\
D\omega ^{+} &=&\alpha \{S_{Z}-c(S_{R}+b)\}\omega ^{0}+\omega ^{1}\alpha
\{S_{R}+b\}  \nonumber \\
&&\omega ^{+}(1-bb^{*})^{-1}\{(1-b^{*}b)D\ln \alpha
+S_{W}-b^{*}(Db+S_{W^{*}})  \nonumber \\
&&-(S_{R}+b)(a-a^{*}b^{*})\}  \nonumber \\
&&+\omega
^{-}(1-bb^{*})^{-1}\{Db+S_{W^{*}}-bS_{W}-(S_{R}+b)(a^{*}-ab)\}\qquad \; 
\nonumber \\
D\omega ^{-} &=&\omega ^{0}\alpha \{b^{*}[S_{Z}-cS_{R}]-c\}+\omega
^{1}\alpha \{1+b^{*}S_{R}\}  \nonumber \\
&&+\omega ^{-}(1-bb^{*})^{-1}\{(1-bb^{*})D\ln \alpha -(a^{*}-ab)-bDb^{*} 
\nonumber \\
&&+b^{*}[S_{W^{*}}-bS_{W}-S_{R}(a^{*}-ab)]\}  \nonumber \\
&&\omega ^{+}(1-bb^{*})^{-1}\{Db^{*}-(a-a^{*}b^{*})  \nonumber \\
&&+b^{*}[S_{W}-S_{W^{*}}b^{*}-S_{R}(a-a^{*}b^{*})]\}  \nonumber
\end{eqnarray}

\begin{eqnarray*}
D\omega ^{1} &=&\omega ^{0}\{Dc+T_{Z}+aS_{Z}-c(aS_{R}+a^{*}+T_{R})\}+\omega
^{1}\{aS_{R}+a^{*}+T_{R}\} \\
&&\omega ^{+}(1-bb^{*})^{-1}\alpha ^{-1}\{T_{W}+c+Da+aS_{W} \\
&&-b^{*}(Da^{*}+aS_{W^{*}}+T_{W^{*}})-(aS_{R}+a^{*}+T_{R})(a-a^{*}b^{*})\} \\
&&\omega ^{-}(1-bb^{*})^{-1}\alpha ^{-1}\{(Da^{*}+aS_{W^{*}}+T_{W^{*}}) \\
&&-(a^{*}-ab)(aS_{R}+a^{*}+T_{R})-b(T_{W}+c+Da+aS_{W})\}
\end{eqnarray*}

\subsection{{\protect\large D}$g=G_{ij}\omega ^{i}\otimes \omega ^{j}$}

The calculation of the $G_{ij},$ defined by D$g=G_{ij}\omega ^{i}\otimes
\omega ^{j}$ begins with 
\[
D\omega ^{i}=A_{j}^{i}\omega ^{j} 
\]
and 
\begin{eqnarray*}
g &=&\eta _{ij}\omega ^{i}\otimes \omega ^{j}, \\
Dg &=&\eta _{ij}D\omega ^{i}\otimes \omega ^{j}+\eta _{ij}\omega ^{i}\otimes
D\omega ^{j}=\{\eta _{jk}A_{i}^{k}+\eta _{ik}A_{j}^{k}\}\omega ^{i}\otimes
\omega ^{j}\equiv G_{ij}\omega ^{i}\otimes \omega ^{j}.
\end{eqnarray*}
Then, by direct substitution, we have

\begin{eqnarray*}
Dg(x^{a},s,s^{*}) &=&2(A_{+}^{0}-A_{1}^{-})\omega ^{(+}\otimes \omega
^{1)}+2(A_{-}^{0}-A_{1}^{+})\omega ^{(-}\otimes \omega ^{1)} \\
&&2A_{0}^{1}\omega ^{0}\otimes \omega ^{0}+2(A_{+}^{1}-A_{0}^{-})\omega
^{(0}\otimes \omega ^{+)} \\
&&+2(A_{-}^{1}-A_{0}^{+})\omega ^{(0}\otimes \omega ^{-)}+2A_{1}^{1}\omega
^{(0}\otimes \omega ^{1)} \\
&&-2(A_{+}^{+}+A_{-}^{-})\omega ^{(+}\otimes \omega ^{-)}-2A_{-}^{+}\omega
^{-}\otimes \omega ^{-}-2A_{+}^{-}\omega ^{+}\otimes \omega ^{+}
\end{eqnarray*}

with

\[
G_{ij}=\left\| 
\begin{array}{lllll}
ij & {\small 0} & {\small +} & {\small -} & {\small 1} \\ 
{\small 0} & 2A_{0}^{1} & (A_{+}^{1}-A_{0}^{-}) & (A_{-}^{1}-A_{0}^{+}) & 
A_{1}^{1} \\ 
{\small +} & (A_{+}^{1}-A_{0}^{-}) & -2A_{+}^{-} & (A_{+}^{+}+A_{-}^{-}) & 
(A_{+}^{0}-A_{1}^{-}) \\ 
{\small -} & (A_{-}^{1}-A_{0}^{+}) & (A_{+}^{+}+A_{-}^{-}) & -2A_{-}^{+} & 
(A_{-}^{0}-A_{1}^{+}) \\ 
{\small 1} & A_{1}^{1} & (A_{+}^{0}-A_{1}^{-}) & (A_{-}^{0}-A_{1}^{+}) & 0
\end{array}
\right\| . 
\]

and explicitly,

$G_{11}=0,$

$G_{-1}=-b(1-bb^{*})^{-1}\alpha ^{-1}-\alpha (S_{R}+b),$

$G_{+1}=(1-bb^{*})^{-1}\alpha ^{-1}-\alpha \{1+b^{*}S_{R}\},$

$G_{--}=-2(1-bb^{*})^{-1}\{Db+S_{W^{*}}-bS_{W}-(S_{R}+b)(a^{*}-ab)\}+\alpha
(S_{R}+b),$

$G_{+-}=(1-bb^{*})^{-1}\{2(1-b^{*}b)D\ln \alpha +(1-bb^{*})S_{W}-D(bb^{*})$

$\qquad \qquad -(S_{R}+b)(a-a^{*}b^{*})-(a^{*}-ab)[1+S_{R}b^{*}]\},$

$G_{++}=-2(1-bb^{*})^{-1}%
\{Db^{*}-(a-a^{*}b^{*})+b^{*}[S_{W}-S_{W^{*}}b^{*}-S_{R}(a-a^{*}b^{*})]\},$

$G_{01}=(aS_{R}+a^{*}+T_{R}),$

$G_{0-}=(1-bb^{*})^{-1}\alpha
^{-1}\{(Da^{*}+aS_{W^{*}}+T_{W^{*}})-(a^{*}-ab)(aS_{R}+a^{*}+T_{R})$

$\qquad \qquad -b(T_{W}+c+Da+aS_{W})\}-\alpha \{S_{Z}-c(S_{R}+b)\},$

$G_{0+}=(1-bb^{*})^{-1}\alpha
^{-1}\{T_{W}+c+Da+aS_{W}-b^{*}(Da^{*}+aS_{W^{*}}+T_{W^{*}})$

$\qquad \qquad -(aS_{R}+a^{*}+T_{R})(a-a^{*}b^{*})\}-\alpha
\{b^{*}[S_{Z}-cS_{R}]-c\},$

$G_{00}=2\{Dc+T_{Z}+aS_{Z}-c(aS_{R}+a^{*}+T_{R})\}.$


\begin{thebibliography}{99}
\bibitem{N}  Though there is a simple heuristic argument for the
four-dimensionality of the solution space, recently, in a private
communication, Peter Vassiliou and Niky Kamran have given a rigorous proof
of this.
\end{thebibliography}
\end{document}